\documentclass[aps,twocolumn,epsf,floatfix]{revtex4}
\usepackage{graphicx}
\usepackage{epsfig}
\begin{document}

\title{Measurement of Two-Qubit States by a Two-Island
Single Electron Transistor}
\author{Tetsufumi Tanamoto$^1$ and Xuedong Hu$^2$}
\affiliation{$^1$Corporate R\&D Center, Toshiba Corporation,
Saiwai-ku, Kawasaki 212-8582, Japan \\
$^2$Department of Physics, University at Buffalo, SUNY, Buffalo,
NY 14260-1500}

\draft
\date{\today}
\begin{abstract}
We solve the master equations of two charged qubits measured by a
single-electron transistor (SET) consisted of two islands.  We show that in
the sequential tunneling regime the SET current can be used for reading out
results of quantum calculations and providing evidences of two-qubit
entanglement, especially when the interaction between the two qubits is weak.  
\end{abstract}
\maketitle

%
Quantum information processing in solid state nanostructures has attracted
wide spread attention because of the potential scalability of such devices. 
Within this context, quantum measurement in mesoscopic systems is a crucial
issue and is being carefully analyzed both experimentally
\cite{Nakamura,Fujisawa,Wiel,Aassime} and theoretically 
\cite{Gurvitz,Goan,Makhlin,Kane,Loss,Korotkov,Tanamoto,Miranowicz},
so that proper measurements can be designed to extract the maximal amount 
of information contained in a solid state qubit (or qubits).
One prominent example is a single-electron transistor (SET), whose current is
particularly sensitive to the charge degrees of freedom through gate potential
variations on its central island(s).  Indeed, with a radio-frequency SET,
electrons can be counted  at frequencies up to 100 MHz \cite{Aassime}, so
that if the states of a qubit can be distinguished by charge locations, an
SET can be used to measure the qubit states.

\begin{figure}
\includegraphics[width=8.5cm]{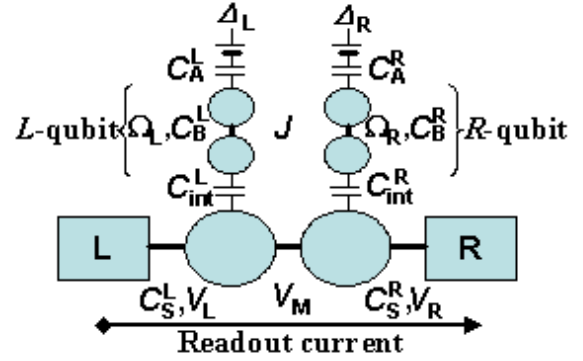}
\caption{Qubits are capacitively coupled to a two-island SET, which acts as
a charge detector. $N(\geq\!2)$ qubits are arranged between source and drain.}
\label{system}
\vspace*{-0.1in}
\end{figure}

Recently, two-qubit coherent evolution and possibly entanglement have been
observed in capacitively coupled Cooper pair boxes \cite{Pashkin}. 
The realization and detection of two-qubit entanglement are crucial
milestones for the study of solid state quantum computing.
In this Letter we study a novel scheme for the quantum measurement of two
charge qubits ($N\!=\!2$), which can be extended to the detection of
moderately larger number of qubits ($N\!>\!2$).
Specifically, the target qubits being constantly measured are double dot
charge qubits \cite{Tanamoto}, whose states are the different spatial
distributions of the excess electron on the double dot.  The quantum detector
is a two-island SET ($N\!=\!2$), with each island coupled to a 
qubit capacitively, as illustrated in Fig.~\ref{system}.  Our objective is to 
demonstrate the capability of this two-island SET in detecting and 
differentiating two-qubit quantum states.  In particular, we develop a master
equation formalism from microscipic Hamiltonian to describe the 
readout current of the SET in its sequential tunneling regime. 
Under the condition that the relaxation time of SET current is sufficiently
long compared to the period of qubit oscillations, 
we clarify three major issues regarding the capability of the
two-island SET layout: whether the
two-qubit eigenstates 
$\{ |00\rangle,
|01\rangle, |10\rangle, {\rm and} |11\rangle\}$ can be distinguished;
whether entangled states and product states can be distinguished;
and whether Zeno effect can be seen in the two qubits.

\begin{figure}
\includegraphics[width=8.5cm]{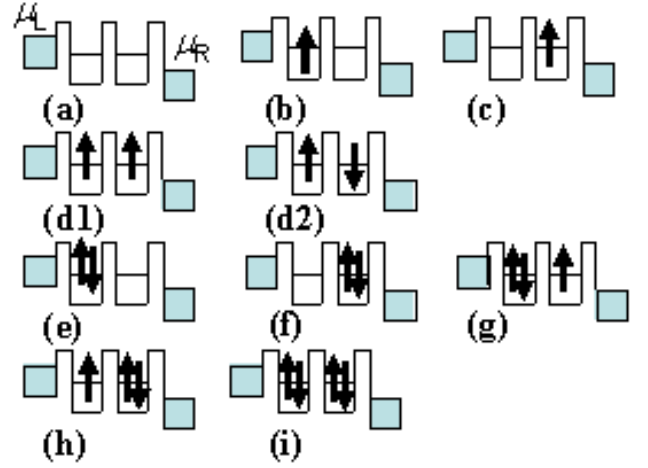}
\caption{Electronic states in the detector.}
\label{SET}
\end{figure}

The Hamiltonian for the combined two qubits and the two-island SET can
be written as follows:
\begin{equation}
H=H_{\rm qb}+H_{\rm set}+H_{\rm int}.
\end{equation}
where $H_{\rm qb}$, $H_{\rm set}$, and $H_{\rm int}$ are the Hamiltonians
of the two qubits, the
SET, and the interaction between the qubits and the SET, respectively.
$H_{\rm qb}$ describes the two interacting (left and right, as illustrated
in Fig.~1) qubits, each consisted of two tunnel-coupled quantum dots
(QDs) and containing one excess charge \cite{Tanamoto}:
\begin{equation}
H_{\rm qb} \!=\! \sum_{\alpha\!=\!L,R}
(\Omega_\alpha \sigma_{\alpha x} \!+\! \Delta_\alpha \sigma_{\alpha z})
\!+\! J \sigma_{Lz} \sigma_{Rz}
\end{equation}
where $\Omega_L (\Omega_R$) and $\Delta_L(t) [\Delta_R(t)]$
are the inter-QD (but intra-qubit) tunnel coupling and energy difference
in the left (right) qubit.  Here we use the spin notation such that
$\sigma_{\alpha x} \equiv a_\alpha^\dagger b_\alpha
+ b_\alpha^\dagger a_\alpha$ and
$\sigma_{\alpha z} \equiv a_\alpha^\dagger a_\alpha
- b_\alpha^\dagger b_\alpha$ ($\alpha=L,R$), where
$a_\alpha$ and $b_\alpha$ are the annihilation operators of an electron
in the upper and lower QDs of each qubit.
$J$ is a coupling constant between the two qubits, originating
from capacitive couplings in the QD system \cite{Tanamoto}.
$|\!\uparrow\rangle$ and $|\!\downarrow\rangle$ refer to the
two single-qubit states in which
the excess charge is localized in the upper and lower dot,
respectively.  $\Delta_\alpha$ ($\alpha=L,R)$ are
bias gate voltages 
applied on the qubits, which can be used to tune the qubit energy
splittings and are used for the manipulation of
these charge qubits during quantum calculations \cite{Tanamoto}.
The SET part of the Hamiltonian $H_{\rm set}$ is written as:
\begin{eqnarray}
H_{\rm set}\!=\!\!\!\!\sum_{\alpha\!=L,R}\!\left[\sum_{i_\alpha,s}
E_{i_\alpha} c_{i_\alpha s}^\dagger c_{i_\alpha s}
\!+\!\sum_{s} E_{d_\alpha s} d_{\alpha s}^\dagger d_{\alpha s}
\!+\! U_\alpha n_{\alpha \uparrow}  n_{\alpha \downarrow} \right] & &
\nonumber \\
& & \hspace*{-3.2in}+\sum_{\alpha\!=\!{\small L,R}} \sum_{i_\alpha,s}\!
V_{\alpha s} \left(c_{i_\alpha s}^\dagger d_{\alpha s}
\!+\! d_{\alpha s}^\dagger c_{i_\alpha s} \right)
\nonumber \\
& & \hspace*{-3.2in}+\sum_{s} V_{Ms} \left(d_{Ls}^\dagger d_{Rs}\! +\!
d_{Rs}^\dagger d_{Ls} \right).
\label{eqn:H_set}
\end{eqnarray}
Here $c_{i_{L}s}$($c_{i_{R}s}$) is the annihilation operator of an electron
in $i_L$th ($i_R$th) level ($i_L(i_R)=1,...,n)$, in the left(right) electrode,
$d_{Ls}$($d_{Rs}$) is the electron annihilation operator of the left (right)
SET island, $s \in
\{\uparrow,\downarrow\}$ is the electron spin, and $n_{\alpha s}\equiv
d_{\alpha s}^\dagger d_{\alpha s}$ is the number of electron on each island.
Here we assume only one energy level on each
island.  $V_{Ls}$($V_{Rs}$) and $V_{Ms}$ are the tunneling 
strength of electrons between left (right) 
electrode and the left (right) island and that between the two islands. 
$U_L(U_R)$ is the on-site Coulomb energy
of double occupancy in the left (right) island.  Finally, the interaction 
between the qubits and the SET, described by $H_{\rm int}$, are
capacitive couplings between the qubits and the two SET islands:
\begin{equation}
H_{\rm int} = \sum_{s} \left( E_{\rm int}^{L}
d_{Ls}^\dagger d_{Ls} \sigma_{Lz}
\!+\!  E_{\rm int}^{R}  d_{Rs}^\dagger d_{Rs} \sigma_{Rz} \right).
\label{eqn:int}
\end{equation}
Consequently, the energy level of an SET island is raised by
$E_{\rm int}^{\alpha}\! \sim e C_{\rm int}^{\alpha}C_{\rm A}^{\alpha}/
C_{\rm S}^{\alpha}/(C_{\rm A}^{\alpha}C_{\rm int}^{\alpha}
\!+\!C_{\rm B}^{\alpha}[C_{\rm A}^{\alpha}\!+\!C_{\rm int}^{\alpha}])$ 
if the charge in the corresponding qubit is located in the lower 
QD \cite{Makhlin}.
%
The electronic states of the qubits also influence the tunneling rates 
$\Gamma^\alpha(E)\!=\!2\pi \rho_\alpha (E)|V_\alpha(E)|^2$ 
($\rho_\alpha$ are densities of states of the electrodes).
If we define $\{|A \rangle \!\equiv\! |\!\downarrow \downarrow \rangle$,
        $|B \rangle \!\equiv\! |\!\downarrow \uparrow \rangle$,
        $|C \rangle \!\equiv\! |\!\uparrow \downarrow \rangle$,
        $|D \rangle \!\equiv\! |\!\uparrow \uparrow \rangle \}$,
the tunneling rates have the relations;
$\Gamma^L_A\!=\!\Gamma^L_B \!<\!\Gamma^L_C\!=\!\Gamma^L_D$ 
and $\Gamma^R_A\!=\!\Gamma^R_C \!<\!\Gamma^R_B\!=\!\Gamma^R_D$.  

Now we can construct the equations of the qubits-SET density matrix 
elements governed by the above-mentioned Hamiltonian at $T\!=\!0$,
following the procedure developed by Gurvitz \cite{Gurvitz}. 
The possible electronic states in the detector are shown in Fig.~\ref{SET}.
The method is applicable as long as the energy-levels of the islands
is inside the chemical potential $\mu_L$ of the left electrode and $\mu_R$ of
the right electrode, and the tunneling rates are much smaller than the
difference $\mu_L-$ $\mu_R$, {\it i.e.} $\mu_L-\mu_R \gg \{\Gamma^L,
\Gamma^R,V_M$\} \cite{Stoof}.  We consider the following two transport 
processes
separately.  The first case is when the double-occupied states are inside the
range of $\mu_L$ and $\mu_R$ and all electronic states in Fig.2 take part in
the tunneling (finite $U$ model).  The second case is when 
double occupancy of electrons [(e)-(i)] is prohibited (infinite $U$ model).
Experimentally, these two cases are interchangeable by tuning
applied island gate voltages \cite{Double}.

The wave function $|\Psi(t)\rangle$ of the qubits-SET system can be expanded
over the electronic states of the qubits
and the island states of the SET shown in Fig.~\ref{SET}.
Assuming that there is no magnetic field and the tunneling is
independent of spin, 
after a lengthy calculation, 
we obtain 352 equations for density matrix elements
$\rho_{u_1u_2}^{z_1z_2}(t)$
($u_1,u_2$ indicate quantum states of the detector (Fig.~\ref{SET}) and
$z_1,z_2=A,B,C,D$ are those of the qubits) as \cite{num_of_eq}:
\begin{eqnarray}
%
\dot{\rho}_{aa}^{AA}\!&=&\!-2\Gamma^L\rho_{aa}^{AA}
\!-\!i\Omega_R (\rho_{aa}^{BA}\!-\!\rho_{aa}^{AB}) 
\!-\!i\Omega_L (\rho_{aa}^{CA}\!-\!\rho_{aa}^{AC}) 
\nonumber \\ & &  
+ \Gamma^R (\rho_{cc\uparrow}^{AA}
\!+\!\rho_{cc\downarrow}^{AA}),
\nonumber \\
%
\dot{\rho}_{aa}^{AB}
\!&=&\!(i[\!-\!J_A\!+\!J_B]
\!-\!2\Gamma^L)\rho_{aa}^{AB}
\!- i\Omega_R (\rho_{aa}^{BB}\!-\!\rho_{aa}^{AA})
\nonumber \\
& &
\!-\!i\Omega_L (\rho_{aa}^{CB}\!-\!\rho_{aa}^{AD}) 
\!+\!\Gamma^{R}(\rho_{cc\uparrow}^{AB}
\!+\!            \rho_{cc\downarrow}^{AB}),
\nonumber \\
%
...... \nonumber \\
%
\dot{\rho}_{ii}^{CD}\!&=&\!
\! 2(i[\!-\!E_{d_L}^C\!-\!E_{d_R}^C\!+\!E_{d_L}^D\!+\!E_{d_R}^D
\!-\!J_C\!+\!J_D]\!-\!\Gamma^{R'})\rho_{ii}^{CD}
\nonumber \\
& &
-i\Omega_R (\rho_{ii}^{DD}\!-\!\rho_{ii}^{CC})
\!-\!i\Omega_L (\rho_{ii}^{AD}\!-\!\rho_{ii}^{CB})  \nonumber \\
& &
+\Gamma^{L'}(\rho_{hh\uparrow}^{CD}
\!+\!\rho_{hh\downarrow}^{CD}),
\label{eqn:master_eq}
\end{eqnarray}
where 
$J_A=\Delta_L\!+\!\Delta_R\!+\!J$,
$J_B=\Delta_L\!-\!\Delta_R\!-\!J$,
$J_C=-\Delta_L\!+\!\Delta_R\!-\!J$,
$J_D=-\Delta_L\!-\!\Delta_R\!+\!J$,
$E_{d_L}^A=E_{d_L}^B=E_{d_L}\!+\!E_{\rm int}^L$,
$E_{d_L}^C=E_{d_L}^D=E_{d_L}\!-\!E_{\rm int}^L$,
$E_{d_R}^A=E_{d_R}^C=E_{d_R}\!+\!E_{\rm int}^R$,
$E_{d_R}^B=E_{d_R}^D=E_{d_R}\!-\!E_{\rm int}^R$.  
$\Gamma^{\alpha'}\!=\!0$ in infinite $U$ model and 
$\Gamma^{\alpha'}\!=\!\Gamma^{\alpha}$ 
in finite $U$ model. 
%
The readout current $I(t)\!=\!e\dot{N}_R(t)$ can then be written as 
\cite{Gurvitz} 
%
\begin{eqnarray}
I(t)\!&=&\!\!\!\! \!\! \!\! \!\!  \sum_{z=A,B,C,D} \!\! \!\!\!\!\! \!\! 
e\{ \Gamma^R [
      \rho_{cc\uparrow}^{zz}
\!+\! \rho_{cc\downarrow}^{zz}
\!+\! \rho_{d_{\uparrow\uparrow}d_{\uparrow\uparrow}}^{zz}\!\!
\!+\! \rho_{d_{\downarrow\uparrow}d_{\downarrow\uparrow}}^{zz}\!\!
\!+\! \rho_{d_{\uparrow\downarrow}d_{\uparrow\downarrow}}^{zz}\!\!
\!+\!\rho_{d_{\downarrow\downarrow}d_{\downarrow\downarrow}}^{zz}\!\!
\nonumber \\
\!&\!+\!&\! 
2\rho_{ff}^{zz} 
\!+\! \rho_{gg\uparrow}^{zz}
\!+\! \rho_{gg\downarrow}^{zz}
%
\!\!+\!\! 2(\rho_{hh\uparrow}^{zz}
\!+\! \rho_{hh\downarrow}^{zz})]
\!+\!2 \Gamma^{R'} \!\rho_{ii}^{zz}  \}.
\end{eqnarray} 
For simplicity we consider two identical qubits,
with $E_{d_L}\!=\!E_{d_R}$ and $E_{\rm int}\equiv E_{\rm int}^L\!=\!E_{\rm 
int}^R$. 

\begin{figure}
\includegraphics[width=8.5cm]{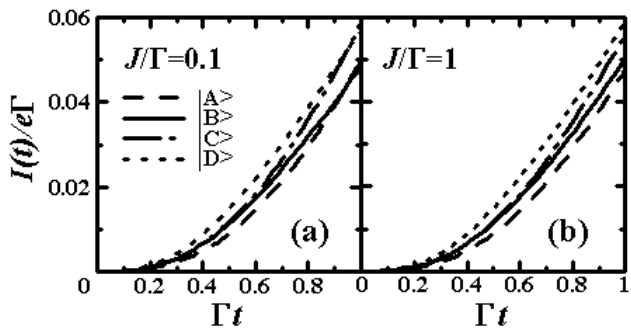}
\caption{
Time dependent readout current characteristics
of the infinite $U$ model for
$|A\rangle=|\downarrow \downarrow\rangle$,
$|B\rangle=|\downarrow \uparrow\rangle$,
$|C\rangle=|\uparrow \downarrow\rangle$,
$|D\rangle=|\uparrow \uparrow\rangle$ as initial states
($t=0$), where
$\Omega_L=\Omega_R=0.75\Gamma$, $V_M=0.5\Gamma$,
$E_{\rm int}^L=E_{\rm int}^R=0.2\Gamma$,
$\Gamma^L_A\!=\!\Gamma^L_B \!=\Gamma^R_A\!=\!\Gamma^R_C=0.8\Gamma$,
$\Gamma^L_C\!=\!\Gamma^L_D=\Gamma^R_B\!=\!\Gamma^R_D=1.2\Gamma$.
(a) $J=0.1\Gamma$, (b) $J=\Gamma$.}
\label{zero-gate-bias}
\end{figure}

We monitor the onset of the readout current to extract
information of the qubit states.
The current begins to flow at $t=0$ and
after a transient region saturates to a steady state value.
In the meantime, the qubits oscillate
with frequencies $\sqrt{\Omega_\alpha^2+\Delta_\alpha^2}$.  
The interaction with the dissipative
current degrades the coherent oscillations and makes the charge distribution
uniform in the qubits at $t\rightarrow \infty$.
Conversely, in the absence of the qubits, the current saturates around
$t \sim \Gamma^{-1}$ where $\Gamma \!\equiv \! \Gamma^L \Gamma^R / 
(\Gamma^L\!+\!\Gamma^R)$, while the qubit charge oscillations modify
the SET current through an effective gate potential on the islands.
Figure \ref{zero-gate-bias} shows the time-dependent current characteristics 
of the infinite $U$ model near $t\sim 0$.  At small $t$ state $|A\rangle$
suppresses the current the most while state $|D\rangle$ the least.
The measurement time $t_m$ that is required to resolve the states of qubits
is estimated as 
$t_{\rm m}^{-1}\!\sim\!{\rm min}\{E_{\rm int}, \Gamma_A^L\!-\!\Gamma_D^L\}$
($\sim 0.5^{-1}\Gamma$ in Fig. \ref{zero-gate-bias}). 
The relative 
magnitude of the current changes after the coherent motions of qubits 
($t>1/\Omega_\alpha$).
Thus the SET current can be used to distinguish the four product states
during $t_{\rm m}<t<1/\Omega_\alpha$.
If the coherent oscillation of the qubits remains 
after $t>\Gamma^{-1}$, as in the present model \cite{phonon},
we can discuss the quantum states of qubits using the steady current formula
($t\rightarrow \infty$) through the SET without the qubits
\cite{Gurvitz}:
\begin{equation}
I_{\rm set}=\frac{e\Gamma V_M^2}
{\epsilon_d^2 \Gamma/(\Gamma^L\!+\!\Gamma^R)\!+\!V_M^2 \!+\! \Gamma^L 
\Gamma^R/4 },
\end{equation}
where $\epsilon_d\equiv E_{d_L}-E_{d_R}$ is the energy difference of the 
two islands.  If $V_M \gg {\Gamma, \epsilon_d, \Omega_\alpha}$, the coupling
between the two islands is strong
and the current mainly reflects the bonding-antibonding state in the detector,
which is not suitable for qubit measurements.
We thus focus on the regime of $V_M < \Omega_\alpha, \Gamma$.
Since $E_{d_L}^A\!-\!E_{d_R}^A\!=\!E_{d_L}^D\!-\!E_{d_R}^D\!=\!0$ and
$E_{d_L}^B\!-\!E_{d_R}^B\!=\!E_{d_R}^C\!-\!E_{d_L}^C\!=\!2E_{\rm int}$, 
the different effects between $|A\rangle$ and $|D\rangle$ and that between
$|B\rangle$ and $|C\rangle$ come from the differences in the tunneling rates.
Moreover, the difference of $|A\rangle$ and $|D\rangle$ from
$|B\rangle$ and $|C\rangle$ becomes obvious in the $E_{\rm int} > V_M$ region.
Thus we call $E_{\rm int} > V_M$ strong measurement regime,
where the four product states can be distinguished,
in contrast to the weak measurement regime of $E_{\rm int} < V_M$.
%

\begin{figure}
\includegraphics[width=8.5cm]{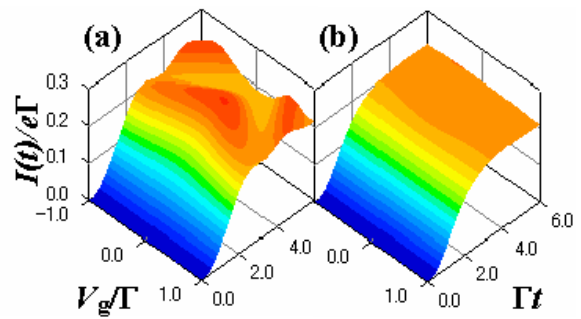}
\caption{
Time dependent readout current characteristics
starting from :(a) $|B\rangle$, (b) singlet state
in the infinite $U$ model for weak measurement case
($E_{\rm int}\!=\!0.2\Gamma<\! V_M\!=\!0.5\Gamma$)
as a function of $V_{\rm g}\!=\!V_{\rm g}^L\!=\!V_{\rm g}^R$.
Parameters are the same as those in Fig.\ref{zero-gate-bias}.
}
\label{weak}
\end{figure}

We can distinguish the current of pure entangled states and that of pure
product states by changing bias voltages $V_g^\alpha=\Delta_\alpha$ in the
regime of $J/\Gamma \ll 1$, where the current depends on the change
of qubit oscillation frequency ($\sim
\sqrt{\Omega_\alpha^2+\Delta_\alpha^2}$).
Figure~\ref{weak}(a) shows the current corresponding to the qubit $|B\rangle$
state in the weak measurement regime of the infinite $U$ model.
We also obtained similar results for the other product states $|A\rangle$,
$|C\rangle$, and $|D\rangle$.
In contrast, the readout current for a two-qubit entangled state
is more uniform compared with the product states
as entangled states generally have less distinct charge distributions.
For example, the density matrix elements for a singlet state
$(|\!\!\uparrow\downarrow\rangle \!-\! |\!\!\downarrow\uparrow
\rangle)/\sqrt{2}=(|C\rangle \!-\! |B\rangle)/\sqrt{2}$ of two free
qubits ($H_{\rm int}=0$) 
satisfy $\dot{\rho}^{BB}\!+\!\dot{\rho}^{CC}\!-\!\dot{\rho}^{BC}\!
-\!\dot{\rho}^{CB}\!=0$ ($\Delta_L=\Delta_R$), which
suggests that entangled states such as the singlet state are less
effective in influencing the readout current.  We believe this
ineffectiveness is related to the fact that logical states encoded
in entangled states are less susceptible to environmental decoherence 
\cite{Palma}.  Indeed, the readout current of this entangled state
is found to be uniform as shown in Fig.~\ref{weak}(b).
We obtained similar results for the other Bell states,
and there is no significant difference between the infinite $U$ model and
the finite $U$ model in the weak measurement regime.

\begin{figure}
\includegraphics[width=8.5cm]{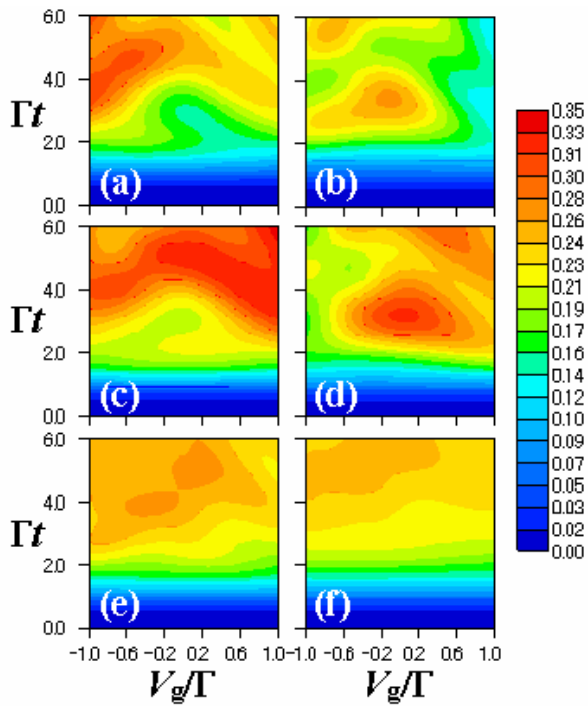}
\caption{
Time dependent readout current characteristics
in the finite $U$ model ($U\!=\!2\Gamma$)for strong measurement case
($E_{\rm int}\!=\!0.8\Gamma >\! V_M\!=\!0.5\Gamma$)
as a function of $V_{\rm g}\!=\!V_{\rm g}^L\!=\!V_{\rm g}^R$.
The initial states are
(a) $|A\rangle$, (b) $|B\rangle$,(c) $|C\rangle$,(d) $|D\rangle$,
(e) triplet state, (f) singlet state.
Parameters other than $E_{\rm int}$ are the same as
those in Fig.\ref{zero-gate-bias}.
}
\label{strong}
\end{figure}

\begin{figure}[t]
\includegraphics[width=8.5cm]{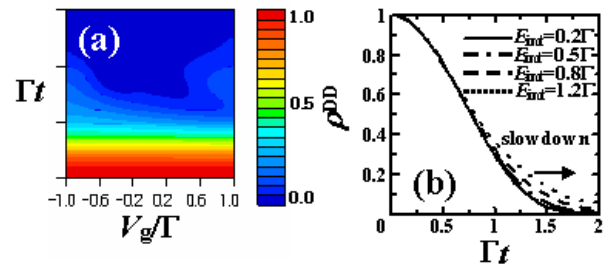}
\caption{
(a) The concurrence of the singlet state.
(b) Example of Zeno effect: oscillation of $\rho^{DD}(t)$ is delayed, where 
the initial state is $|D\rangle$ state ($\rho^{DD}(0)\!=\!1$). Similar effects 
can be seen in other initial states. 
Parameters are the same as those in Fig.\ref{strong}.
}
\label{concurrence}
\end{figure}

In the strong measurement regime ($E_{\rm int} >V_M$),
the current is more sensitive to the charge distributions in the qubits,
and there are differences between the infinite $U$ model and
finite $U$ model. 
We can distinguish the four products more easily through the SET current,
as shown in Fig.~\ref{strong}(a)-(d).
However, currents for the entangled states in the infinite $U$ model
show several similar peaks that reflect the qubit oscillations
and cannot be easily distinguished from the product states.
On the other hand, the finite $U$ model shows distinct uniform structure
compared with
the current of the product states [Fig.~\ref{strong}(e) and (f)].
This shows that, in the finite $U$ model,
redistribution of the electrons through the two islands of the detector 
is energetically favorable under the rather uniform
electric field generated by the entangled qubits.
Figure~\ref{concurrence}(a) shows that the concurrence
(a measure of entanglement \cite{Wootters}
derived from reduced density matrix of two qubits
after tracing over the detector components) of the two qubits
disappears quickly in the cases of strong measurement.
While the coherence quickly degrades, we can see the emergence of the Zeno
effect, in which a continuous measurement slows down transitions
between quantum states due to the collapse of the wavefunctions into observed
states \cite{Gurvitz,Korotkov}.  For instance, Fig.~\ref{concurrence} (b)
shows that, as $E_{\rm int}$ increases, the oscillations of density matrix
elements of the qubits (e.g. $\rho^{DD}$) are delayed, which is a clear
evidence of the slowdown described by the Zeno effect in the two qubits. 

Our numerical results above are applicable to a wide range of pure
product and entangled states.  For example, in the
entangled states
$\cos\theta |\!\uparrow\downarrow\rangle
+e^{i\varphi} \sin\theta |\!\uparrow\downarrow\nobreak\rangle$,
we found that the uniformity of the readout current holds
approximately up to $|\theta \pm \pi/4 |,|\varphi| < \pi /12$.
The pure entangled states are more robust beyond 
the spatial distribution of the wave functions. 
Although the product states 
$\!\prod_{\alpha\!=\!L,R}
[\cos (\frac{\theta_\alpha}{2}) {\rm e}^{-i\frac{\varphi_\alpha}{2}} 
|\uparrow \rangle_\alpha
\!+\! \sin (\frac{\theta_\alpha}{2}) {\rm e}^{i\frac{\varphi_\alpha}{2}} 
|\downarrow \rangle_\alpha
]$
seem to have similarly uniform wave functions when $\theta_L=\pm \theta_R$ and
$\varphi_L =\pm \varphi_R=0,\pi$ (compared to the entangled states mentioned
above), the corresponding currents reflect the coherent oscillations
of the qubits when the gate bias changes between $V_g^L=V_g^R$ and
$V_g^L=-V_g^R$.

%
Since the detection scheme discussed here is based on measuring small
current differences in the transient regime, it is important to analyze
whether the present day technology can achieve the necessary
sensitivity.  The state of the art technology allows the measurement of
1 pA current with dynamics in the GHz frequency range with repeated
measurement techniques \cite{Nakamura,Fujisawa,Gardelis}.  According to our
Figs.~3-5, our scheme requires measuring a 0.1 pA current that changes
in the nanosecond time scale (assuming a $\Gamma$ in the order of 100
MHz, a reasonable figure because $E_{\rm int}$ would be in the order of
100 MHz if all capacitances are 100 aF), which is at the edge of the
current measurement technology.  Thus, with a similar design of repeated
measurement \cite{Nakamura,Fujisawa,Gardelis}, our detection scheme should be
experimentally feasible in the near future.

In conclusion, we have solved master equations and described various
time-dependent measurement processes of two charge qubits.
The current through the two-island SET is shown to be an effective
means to measure results of quantum calculations and entangled states. 

We acknowledge discussions with N. Fukushima, S. Fujita, and M. Ueda. 



%


\begin{thebibliography}{99}
\vspace*{-0.2in}
\bibitem{Nakamura}
Y. Nakamura {\it et al.}, Phys. Rev. Lett. {\bf 88} 047901 (2002). 

\bibitem{Fujisawa}
T. Fujisawa {\it et al.}, Phys. Rev. Lett. {\bf 88}, 236802 (2002).
Phys. Rev. B {\bf 63} 081304 (2001).

\bibitem{Wiel}
W. G. van der Wiel {\it et al.}, Rev. Mod. Phys. {\bf 75} 1 (2003).

\bibitem{Aassime}
A. Aassime {\it et al.}, Phys. Rev. Lett. {\bf 86}, 3376 (2001). 
Schoelkopf {\it et al.}, Scicence {\bf 280}, 1238 (1998).

\bibitem{Gurvitz}
S. A. Gurvitz and Ya. S. Prager, Phys. Rev. B {\bf 53}, 15932 (1996).
%
%
B. Elattari and S. A. Gurvitz, Phys. Rev. Lett. {\bf 84}, 2047 (2000).
%

\bibitem{Goan}
H. S. Goan, {\it et al.} 
Phys. Rev. B {\bf 63}, 125326 (2001).
%

\bibitem{Makhlin}
Y. Makhlin {\it et al.}, 
Rev. Mod. Phys. {\bf 73}, 357 (2001).
%

\bibitem{Kane}
B. E. Kane {\it et al.},  
Phys. Rev. B {\bf 61}, 2961 (2000).
 
\bibitem{Loss}
D. Loss and E.V. Sukhorukov, Phys. Rev. Lett. {\bf 84}, 1035 (2000).
%

\bibitem{Korotkov}
A.N. Korotkov, Phys. Rev. B {\bf 60}, 5737 (1999); 
Phys. Rev. A {\bf 65}, 052304 (2002).
%


\bibitem{Tanamoto}
T. Tanamoto, Phys. Rev. A {\bf 64}, 062306 (2001); {\it ibid}
{\bf 61}, 022305 (2000).

\bibitem{Miranowicz}
A. Miranowicz {\it et al.}, Phys. Rev. A {\bf 65} 062321 (2002);
Y. Liu {\it et al.}, 
Phys. Rev. A {\bf 65}, 042326 (2002).



\bibitem{Pashkin} Yu.A. Pashkin {\it et al.},
Nature {\bf 421}, 823 (2003).


\bibitem{Stoof}
T. H. Stoof and Yu. V. Nazarov, Phys. Rev. B {\bf 53} 1050 (1996).

\bibitem{Double} One limitation of the present formulation is that we cannot
treat the boundary region where the energy of a double occupied island equals
the Fermi energy of an electrode.

\bibitem{num_of_eq} 
We included $(u_1,u_2)=\{((a,a)$, $(b,b)$, $(c,c)$, $(b,c)$, $(d_1,d_1)$, 
$(d_2,d_2)$, $(d_1,d_2)$,
$(e,e)$, $(f,f)$, $(e,d2)$, $(f,d2)$, 
$(e,f)$, $(g,g)$, 
$(h,h)$, $(g,h)$, $(i,i)\}$ where each 
has real and imaginary parts.

\bibitem{phonon}
We ignore other origins of decoherence, such as phonons, or trapped charges
that 
generate the $1/f$ fluctuations.

\bibitem{Palma} G.M. Palma {\it et al.}, Proc. R. Soc. Lond. {\bf 452}, 567
(1996); P. Zanardi, Phys. Rev. A {\bf 57}, 3276 (1998); D. Lidar
{\it et al.}, Phys. Rev. Lett. {\bf 81}, 2594 (1998).

\bibitem{Wootters}
W. K. Wootters, Phys. Rev. Lett. {\bf 80}, 2245 (1998).

\bibitem{Gardelis}
S. Gardelis {\it et al}., Phys. Rev. B {\bf 67}, 073302 (2003) 

\end{thebibliography}
\end{document}